\documentclass[superscriptaddress,twocolumn,showpacs,
preprintnumbers,amsmath,amssymb,nofootinbib,
longbibliography,aps,prd,10pt]{revtex4-1}
\usepackage{graphicx}

\usepackage{color}

\usepackage{soul}
\usepackage{bbold}
\usepackage{ulem} 

\usepackage{amsmath}
\usepackage{amssymb}
\usepackage{amsfonts}
\usepackage[usenames,dvipsnames]{xcolor}
\usepackage{graphicx}  
\usepackage{enumerate}

\usepackage[colorlinks=true,
			citecolor=green,
			linkcolor=red,
			filecolor=cyan,
			urlcolor=magenta,
			backref=false]{hyperref}

\newcommand{\hh}{,\hspace{0.5cm}}
\newcommand{\hhh}{,\hspace{0.2cm}}

\newcommand{\ts}[1]{{\boldsymbol{#1}}}         

\newcommand{\dd}{\mbox{d}}

\newcommand{\be}{\begin{equation}}             
\newcommand{\ee}{\end{equation}}               
\newcommand{\ba}{\begin{eqnarray}}             
\newcommand{\ea}{\end{eqnarray}}               
\newcommand{\n}[1]{\label{#1}}

\begin{document}

\title{Principal Killing strings in higher-dimensional Kerr--NUT--(A)dS spacetimes}

\author{Jens Boos}

\email{boos@ualberta.ca}

\author{Valeri P. Frolov}

\email{vfrolov@ualberta.ca}

\affiliation{Theoretical Physics Institute, University of Alberta, Edmonton, Alberta, Canada T6G 2E1}

\date{\today}

\begin{abstract}
We construct special solutions of the Nambu--Goto equations for stationary strings in a general Kerr--NUT--(A)dS spacetime in any number of dimensions. This construction is based on the existence of explicit and hidden symmetries generated by the principal tensor which exists for these metrics. The characteristic property of these string configurations, which we call ``principal Killing strings,'' is that they are stretched out from ``infinity'' to the horizon of the Kerr--NUT--(A)dS black hole and remain regular at the latter. We also demonstrate that principal Killing strings extract angular momentum from higher-dimensional rotating black holes and interpret this as the action of an asymptotic torque.
\end{abstract}

\pacs{04.20.Jb, 04.50.Gh, 04.70.Bw, 11.27.+d \hfill Alberta-Thy-14-17}

\maketitle

\section{Introduction}

The interaction of cosmic strings with black holes is a subject which is interesting for several reasons. Both objects are non-local and relativistic and there are many interesting physical effects which occur when these objects interact (see e.g.~\cite{anderson2015mathematical}). Moreover, cosmic strings can be used for mining energy and angular momentum from black holes \cite{Lawrence:1993sg,Frolov:2000kx,PhysRevD.54.5093,0264-9381-14-5-015,1402-4896-62-2-3-005,Kinoshita:2016lqd}. The induced metric on the worldsheet of a cosmic string which pierces the black hole contains a two-dimensional horizon, and thus it plays the role of a two-dimensional black hole for the string perturbations which  propagate along the string \cite{PhysRevD.54.5093}. Certainly, the status of these two classes of objects in astrophysics is quite different. We have strong evidence of existences of stellar mass and supermassive black holes. They are rather common objects which manifest their existence in observations, including the remarkable discovery of binary black-hole coalescence, registered by gravitational observatories (by LIGO, and more recently by VIRGO). On the other hand, there are no direct confirmations that topological defects, such as cosmic strings---the existence of which is expected in different theories with topological phase transitions---do really exist in the Universe. Observations of the cosmological microwave background (CMB) allow one to restrict the abundance of heavy (GUT) cosmic strings. Namely, their effect on the CMB is less than 10\% \cite{Ade:2013xla}. However, lighter cosmic strings still may be formed in the corresponding symmetry breaking phase transitions. The search of possible effects of the interaction of such strings with black holes remains an interesting physical and astrophysical problem.

There exists another aspect of this problem. Namely, stationary black holes possess very rich geometrical structures, which allow one to solve many problems. The origin of the explicit and hidden symmetries of stationary black holes is connected with the existence of a so-called principal tensor. This is a rank two closed conformal Killing--Yano form. This object generates a set (tower) of Killing vectors and tensors that is sufficient for the complete integrability of geodesic equations and complete separation of variables in the Kerr metric (historical discussion and references can be found in the recent paper in \emph{Living Reviews in Relativity} \cite{Frolov:2017kze}). A long time ago it was demonstrated that stationary string equations can also be integrated in the Kerr metric \cite{Frolov:1988zn} (as well as its Kerr--(A)dS generalizations \cite{Carter:1989bs}). This result was generalized to stationary strings in the general Kerr--NUT--(A)dS metric in any number of spacetime dimensions \cite{Kubiznak:2007ca}.

Among stationary string solutions there exist special string configurations that penetrate the black hole ergosphere and remain regular at the future event horizon. They are called principal Killing strings \cite{Frolov:2004qw}. Their characteristic property is that their two-dimensional worldsheet is tangent to the Killing vector which is timelike at infinity, and to the principal null geodesics of the Kerr \cite{PhysRevD.54.5093} and Myers--Perry \cite{Frolov:2004qw} metrics. For examples of principal Killing strings in lower dimensions, see \cite{Frolov:1995qp,Frolov:1995vp,Frolov:1996be}. These solutions were generalized to four-dimensional Kerr--NUT--(A)dS spacetimes in our previous work \cite{Boos:2017pyd}.

In this paper we study stationary strings in a general spacetime that admits a principal tensor \cite{Frolov:2017kze}. We do not assume that its metric is a solution of the vacuum Einstein equations. In $(D=2n+\epsilon)$ dimensions, where $\epsilon=0,1$, such a general metric with a principal tensor contains $n$ arbitrary functions of one variables. The class of Kerr--NUT--(A)dS metrics, which form a subset of these so-called off-shell metrics, are obtained by imposing vacuum Einstein equations with cosmological constant. This reduces the arbitrary functions in the off-shell metric to polynomials. The coefficients of these polynomials contain $2n-1$ free parameters that are related to the mass mass, rotation, and NUT parameters of the corresponding black hole. In the present paper we construct principal Killing string solutions in a general off-shell metric using the principal tensor. This construction does not employ a special concrete form of this metric. In this sense our approach is quite different from and more general than the one adopted in \cite{Kubiznak:2007ca,Frolov:2004qw}. Therein, the complete integrability of stationary string equations in Kerr--NUT-(A)dS geometries \cite{Kubiznak:2007ca} implies that a string configuration can be found by quadratures, that is, by a finite number of steps involving algebraic operations and integration. However, even in four dimensions the corresponding integrals are elliptical and in order to obtain an explicit expression for the string configuration one needs to invert these functions. In higher dimensions, the corresponding integrals contain square roots of polynomials of the order higher than 4. That is why in a general Kerr--NUT--(A)dS geometry the solutions for stationary string equations cannot be written in explicit form, but rather presented in a parametrically as a set of equations containing these complicated integrals.

In the present work, we focus on special stationary solutions, which are called principal Killing strings. These solutions can be constructed by using the properties of the principal tensor in the underlying geometry without using the concrete form of the metric.

This paper is organized as follows. In Sec.~II we consider a spacetime which admits a principal tensor\footnote{For definition and properties of the principal tensor see \cite{Frolov:2017kze}.} and discuss the properties of principal null congruences. Such a congruence is formed by null geodesics generated by null eigenvectors of the principal tensor. In Sec.~III we demonstrate that a two-dimensional surface, tangential to principal null geodesics and the (primary) Killing vector, is minimal, and hence it is a solution of the Nambu--Goto equations representing a stationary string. We also obtain the explicit form of such a solution in the Kerr--NUT--(A)dS spacetime with any number of dimensions. To illustrate the properties of principal Killing strings we discuss these solutions explicitly in the Myers--Perry geometry (Sec.~IV). For these metrics the cosmological constant and NUT-parameters vanish and the spacetime is asymptotically flat. The principal Killing string stretches out from ``infinity'' to the horizon and remains regular at the latter. We reproduce the results for the energy and angular momentum transfer through horizon generated by the string, see Ref.~\cite{Frolov:2004qw}, demonstrate the consistency of the calculations of these quantities at the horizon and infinity, and prove a special version of the energy and angular momentum conservation theorem. We also discuss the backreaction problem, and relate (in Sec.~V) the spinning down effect (produced by the angular momentum flux) to the torque applied to the string at infinity. In a general case, the torque and force produced by the string not only lead to a precession of its total angular momentum, but also result in the change of the linear momentum of the black hole. We show how to choose a stationary configuration of several segments of the principal Killing strings that pierce the black hole in such a way that the induced precession and momentum vanish. In Sec.~VI we briefly discuss the obtained results.

\section{Principal tensor and principal null congruencies}

\subsection{Spacetime with a principal tensor}
Let us consider  $D=2n+\epsilon$ dimensional spacetime ($\epsilon=0,1$) which admits a principal tensor $\ts{h}$ \cite{Frolov:2017kze}. Such a tensor is a non-degenerate closed conformal Killing--Yano 2-form that obeys the equation
\be\n{pt}
\nabla_c h_{ab}=g_{ca}\xi_b-g_{cb}\xi_a\hh \xi_a={1\over D-1}\nabla^b h_{ba}\, .
\ee
For such a tensor there exists an orthonormal Darboux frame $(\ts{l}_+,\ts{l}_-,\ts{e}^{\mu},\hat{\ts{e}}^{\mu},\hat{\ts{e}}^0)$ in which one has
\ba
\ts{h}&=&r \ts{l}_+\wedge \ts{l}_- +\sum \limits_{\mu=1}^{n-1} x_{\mu} \ts{e}^{\mu}\wedge \hat{\ts{e}}^{\mu} \, ,\\
\ts{g}&=&-\ts{l}_+ \ts{l}_- - \ts{l}_- \ts{l}_+  +\sum \limits_{\mu=1}^{n-1} (\ts{e}^{\mu}\ts{e}^{\mu}+ \hat{\ts{e}}^{\mu}\hat{\ts{e}}^{\mu})+\epsilon \hat{\ts{e}}^{0}\hat{\ts{e}}^{0}\, .\n{gg}
\ea
Here and later the index $\mu$ takes values $1,\ldots,n-1$. The condition that the principal tensor is non-degenerate means that there are exactly $n$ non-vanishing ``eigenvalues'' $(r,x_{\mu})$ that are functionally independent in some domain. We also assume that in this domain none of the gradients $\nabla_a r$ and $\nabla_a x_{\mu}$ is a null vector.

One can show that $\ts{\xi}$ is a Killing vector, which is called primary and the following relations are valid \cite{Frolov:2017kze}
\be\n{Lgh}
{\cal L}_{\ts{\xi}}\ts{g}={\cal L}_{\ts{\xi}}\ts{h}=0\, .
\ee

The elements of the Darboux basis obey the following normalization conditions
\be
(\ts{l}_+,\ts{l}_-)=-1\hhh (\ts{e}^{\mu},\ts{e}^{\mu})=(\hat{\ts{e}}^{\mu},\hat{\ts{e}}^{\mu})=(\hat{\ts{e}}^{0},\hat{\ts{e}}^{0})=1\, .
\ee
All other scalar products vanish. The basis vectors also obey the following relations
\be\n{lee}
\ts{h}\cdot \ts{l}_{\pm} =\mp r \ts{l}_{\pm} \hhh
\ts{h}\cdot \ts{e}^{\mu} =-x_{\mu} \hat{\ts{e}}^{\mu}\hhh
\ts{h}\cdot  \hat{\ts{e}}^{\mu} =x_{\mu}\ts{e}^{\mu}\hhh
\ts{h}\cdot  \hat{\ts{e}}^{0} =0
\, .
\ee

Sometimes, instead of unit vectors $\ts{e}^{\mu}$ and $\hat{\ts{e}}^{\mu}$ it is more convenient to use their linear combinations
\be
\ts{m}^{\mu}={1\over \sqrt{2}}(\ts{e}^{\mu}+i\hat{\ts{e}}^{\mu})\hh
\bar{\ts{m}}^{\mu}={1\over \sqrt{2}}(\ts{e}^{\mu}-i\hat{\ts{e}}^{\mu})\, .
\ee
They satisfy the following relations:
\ba
&& (\ts{m}^{\mu},\ts{m}^{\mu})=(\bar{\ts{m}}^{\mu},\bar{\ts{m}}^{\mu})=0\hh
(\ts{m}^{\mu},\bar{\ts{m}}^{\mu})=1\, ,\\
&&\ts{h}\cdot \ts{m}^{\mu}=ix_{\mu}\ts{m}^{\mu}\, .\n{hmx}
\ea

In a spacetime with the principal tensor the following results are valid (see \cite{Frolov:2017kze} and references therein):
\begin{itemize}
\item $\ts{h}^{(j)}={1\over j!}\ts{h}^{\wedge j}$ is a closed conformal Killing--Yano $2j$-form;
\item $\ts{f}^{(j)}=* \ts{h}^{(j)}$ is a Killing--Yano $(D-2j)$-form;
\item $k_{(j)}^{ab}={1\over (D-2j-1)!} f^{(j)a}_{\ \ \ c_1...c_{D-2j-1}}f^{(j)bc_1...c_{D-2j-1}}$ is a rank 2 Killing tensor. $\ts{k}_{(0)}=\ts{g}$. In odd dimensions $\ts{f}^{(n)}\sim \partial_{\psi_n}$ is a Killing vector;
\item $\ts{\zeta}_{(j)}=\ts{k}_{(j)}\cdot \ts{\xi} $ $(j=0,\ldots,n-1+\epsilon \equiv m)$ are commuting Killing vectors. For $j=0$ one has $\zeta_{(0)}=\ts{\xi}$, while for $j>0$ they are called secondary Killing vectors;
\item One has ${\cal L}_{\ts{\zeta}_{(j)}}\ts{h}=0$;
\item According to the Frobenius theorem, $n+\epsilon$ commuting Killing vectors are tangent to $(n+\epsilon)$ dimensional submanifolds and there exist such coordinates $\psi_j$ on these Killing submanifolds in which $\ts{\zeta}_{(j)}=\partial_{\psi_j}$;
\item $\psi_0 \equiv \tau$ is the timelike Killing coordinate while  $\psi_k$ ($k=1,\dots,n-1+\epsilon \equiv m$) are the spacelike azimuthal Killing coordinates.
\item The eigenvalues $(x_{\mu},r)$, $(\mu=1,\ldots,n-1)$, are constant on the Killing submanifolds and together with $\psi_j$ they form a coordinate chart on the spacetime. $x_{\mu}$ and $r$ are called polar and radial coordinates, respectively.
\end{itemize}

The last property can be easily proven by using the relation ${\cal L}_{\ts{\zeta}_{(j)}}\ts{h}=0$. Denote by $\ts{v}$ any of the eigenvectors $(\ts{l}_{\pm},\ts{m}^{\mu},\bar{\ts{m}}^{\mu}, \ts{e}^0)$ of the principal tensor and by $y$ its eigenvalue such that
\be\n{hvy}
\ts{h}\cdot \ts{v}=y\ts{v}\, .
\ee
We also denote $\ts{V}={\cal L}_{\ts{\zeta}_{(j)}}\ts{v}$. Then (\ref{hvy}) implies
\be
\ts{h}\cdot \ts{V}=y\ts{V}+ (\partial_{\psi_j}y) \ts{v}\, .
\ee
This relation is valid only if $\ts{V}=C\ts{v}$ and $\partial_{\psi_j}y=0$. In other words, $x^{\mu}$ and $r$ are constant on the Killing submanifolds.

\subsection{Principal null geodesics}

The null eigenvectors $\ts{l}_{\pm}$ of the principal tensor $\ts{h}$ are called principal null vectors\footnote{
Such a vector satisfies the relation \cite{Frolov:2017kze,Ortaggio:2017ydo}
\[
{l}_{[e}C_{a]b[cd}l_{f]}l^b=0 \, ,
\]
and hence it defines a multiple Weyl aligned null direction.
}. They possess a number of remarkable properties. In this section we collect some of them.

First, let us demonstrate that $\ts{l}_{\pm}$ are tangent vectors to null geodesics. We denote
\be
\ts{w}_{\pm}=\nabla_{\ts{l}_{\pm}}\ts{l}_{\pm}\, ,
\ee
where $\nabla_{\ts{l}_{\pm}} \equiv l^c_{\pm}\nabla_c$. Using (\ref{pt}) one gets
\be
\nabla_{\ts{l}_{\pm}}h_{ab}=l_{\pm a}\xi_b-l_{\pm b}\xi_a \, .
\ee
Applying $\nabla_{\ts{l}_{\pm}}$ to the first of the relations in (\ref{lee}) one finds
\be\n{hw}
\ts{h}\cdot \ts{w}_{\pm}\pm r \ts{w}_{\pm}=-[(\ts{\xi},\ts{l}_{\pm})\pm \nabla_{\ts{l}_{\pm}} r]\ts{l}_{\pm}\, .
\ee
Let us write $\ts{w}_{\pm}$ in the form
\be
\ts{w}_{\pm}=C_{\pm}\ts{l}_{\pm}+\ts{p}_{\pm}\, ,
\ee
where $\ts{p}_{\pm}$ is a linear combination of $\ts{l}_{\mp}$, $\ts{e}^{\mu}$, $\hat{\ts{e}}^{\mu}$ and (in the case of odd dimensions) $\hat{\ts{e}}^{0}$. It is easy to check that Eq.~\eqref{hw} can be satisfied only when $\ts{p}_{\pm}=0$, such that
\be
\nabla_{\ts{l}_{\pm}}\ts{l}_{\pm}=C_{\pm}\ts{l}_{\pm}\hhh (\ts{\xi},\ts{l}_{\pm})\pm \nabla_{\ts{l}_{\pm}} r=0\, .
\ee
The first of these relations demonstrates that the integral curves of $\ts{l}_{\pm}$ are null geodesics. Since (\ref{lee}) determines $\ts{l}_{\pm}$ up to a rescaling one can use this transformation so that the vectors
\be
\ts{\ell}_{\pm} \equiv \beta_{\pm}\ts{l}_{\pm}
\ee
are tangent to null geodesics in the affine parametrization. However, the normalization condition for these null vectors becomes
\be
(\ts{\ell}_+,\ts{\ell}_-)=-\beta_+ \beta_-\, .
\ee

It is easy to check that
\be
\nabla_{\ts{l}_{\pm}}(\ts{\xi},\ts{l}_{\pm})=(\ts{\xi},\ts{w}_{\pm})\, ,
\ee
such that $(\ts{\xi},\ts{\ell}_{\pm})$ is constant along the null geodesic in the affine parametrization. We assume that $\ts{\ell}_{\pm}$ are future-directed. If the primary Killing vector is timelike in some domain, we choose
\be
(\ts{\xi},\ts{\ell}_{\pm})=-1\, ,
\ee
which immediately implies
\be\n{lrr}
\nabla_{\ts{\ell}_{\pm}} r=\pm 1\, .
\ee

Using a similar method we can prove that
\be\n{nlx}
\nabla_{\ts{l}_{\pm}}x_{\mu}=0\, .
\ee
For this purpose we denote $\ts{U}^{\mu} \equiv \nabla_{\ts{l}_{\pm}}\ts{m}^{\mu}$. Applying the operator $\nabla_{\ts{l}_{\pm}}$ to (\ref{hmx}) one finds
\be
\ts{h}\cdot \ts{U}^{\mu}-ix_{\mu}\ts{U}^{\mu}=i (\nabla_{\ts{l}_{\pm}} x_{\mu}) \ts{m}^\mu -(\ts{\xi},\ts{m}^{\mu}) \ts{l}_{\pm}\, .
\ee
Contracting this relation with $\bar{\ts{m}}^{\mu}$ yields
\be\n{UUU}
i \nabla_{\ts{l}_{\pm}} x_{\mu}=(\bar{\ts{m}}^{\mu},\ts{h}\cdot \ts{U}^{\mu})-ix_{\mu}(\bar{\ts{m}}^{\mu},\ts{U}^{\mu})\, .
\ee
Let us write $\ts{U}^{\mu}$ in the form
\be
\ts{U}^{\mu}=C \ts{m}^{\mu}+\ts{q}\, .
\ee
It is easy to see that the right-hand side of (\ref{UUU}) vanishes, and hence (\ref{nlx}) is valid. Finally, let us note that for affinely parametrized principal null geodesics the complete set of integrals of motion is given by \cite{Kubiznak:2008zs}
\be
\kappa_{j}\equiv k_{(j) ab} \ell^a \ell^b=0\hhh \Psi_j\equiv\zeta_{(j)a} \ell^a=-A_{n}^{(j)}\hhh\Psi_n=0\, ,
\ee
where we defined
\be
A^{(k)}_\mu = \sum\limits_{\substack{\nu_1,\dots,\nu_k=1 \\ \nu_1 < \dots < \nu_k \\\nu_i \not= \mu }}^n x{}_{\nu_1}^2 \dots \, x{}_{\nu_k}^2\, .
\ee

\section{Principal Killing strings}
\label{sec:pks}

\subsection{Principal Killing surfaces}

We can now define a principal Killing string. Let us consider a two-dimensional surface $\Sigma_{\pm}$ which is tangent to the primary Killing vector $\ts{\xi}$ and the one of the  principal null vectors $\ts{\ell}_{\pm}$. In order to prove that such a surface exists we show first that the vectors $\ts{\xi}$ and $\ts{\ell}_{\pm}$ commute
\be\n{xll}
[\ts{\xi},\ts{\ell}_{\pm}]=0\, .
\ee
After this we prove that $\Sigma_{\pm}$ is a minimal surface, and hence it is a solution of the string equation of motion representing the worldsheet of the special stationary string configuration.

Let us start with the first property. Let us apply ${\cal L}_{\ts{\xi}}$ to the relation
$\ts{h}\cdot \ts{\ell}_{\pm} =\mp r \ts{\ell}_{\pm}$. Then, since ${\cal L}_{\ts{\xi}}\ts{h}={\cal L}_{\ts{\xi}} r=0$, one has
\be
\ts{h}\cdot {\cal L}_{\ts{\xi}}\ts{\ell}_{\pm} =\mp r {\cal L}_{\ts{\xi}}\ts{\ell}_{\pm}\, .
\ee
This relation implies
\be
{\cal L}_{\ts{\xi}}\ts{\ell}_{\pm}=C \ts{\ell}_{\pm}\, .
\ee
Applying ${\cal L}_{\ts{\xi}}$  to the relation $\ts{\xi}\cdot \ts{\ell}_{\pm}=\pm 1$ one gets  $\ts{\xi}\cdot {\cal L}_{\ts{\xi}}\ts{\ell}_{\pm}=0$. Hence $C=0$ and ${\cal L}_{\ts{\xi}}\ts{\ell}_{\pm}\equiv [\ts{\xi},\ts{\ell}_{\pm}]=0$.

To prove that the principal Killing surface is minimal we shall need the following result:
The principal null vectors $\ts{\ell}_{\pm}$ are eigenvectors of 2-form $F_{ab}= \nabla_a\xi_b$:
\be\n{Flk}
F{}^a{}_b \ell{}^b_{\pm} = \kappa_{\pm} \ell{}^a_{\pm}\, .
\ee
Multiplying relation (\ref{pt}) by $\xi^c$ one obtains $\xi{}^c \nabla{}_c h{}_{ab} = 0$. Thus the equation ${\cal L}_{\ts{\xi}}\ts{h}=0$, see Eq.~\eqref{Lgh}, takes the form
\be
F_a{}^b h{}_{bc}= F{}_{c}{}^b h{}_{ba}\, .
\ee
Then, denoting $V{}^a = F{}^a{}_b \ell{}_{\pm}^b$, one has
\begin{align}
h{}^a_b V{}^b = h{}^a_b F{}^b{}_c \ell{}_{\pm}^c = F{}^a{}_b h{}^b{}_c \ell_{\pm}^c =\mp r V{}^a .
\end{align}
Since $\ts{h}$ is non-degenerate, all its eigenvalues have algebraic multiplicity of 1, and hence $V{}^a \sim \ell^a_\pm$, so that \eqref{Flk} is valid. Multiplying \eqref{Flk} by $\ts{\xi}$ one obtains
\be
\kappa_{\pm}={1\over 2}\ell_{\pm}^a (\ts{\xi}^2)_{;a}\, .
\ee

\subsection{Principal Killing surfaces are minimal}

Equation \eqref{xll} and the Frobenius theorem allow us to define the coordinates $(z^A,y^i)$ in which the equation of the Killing surface $\Sigma_{\pm}$ takes the form $y^i=\text{const}$, while $z^A=(v,\lambda_{\pm})$ are coordinates on $\Sigma_{\pm}$ such that
\begin{align}
\ts{\xi} = \partial{}_v , \quad \ts{\ell}_\pm = \partial{}_{\lambda_{\pm}} .
\end{align}
Let us define functions $Y^a(z^A,y^i)$, $A=0,1$ and $i=2,\ldots,D-2$, such that for fixed values $y^i$ they determine the embedding of the two-surface $\Sigma_{\pm}$ in the bulk spacetime and
\be\n{YYY}
Y^a{}_{,v}=\xi^a\hh Y^a{}_{,\lambda_{\pm}}=\ell_{\pm}^a\, .
\ee

The induced metric and its inverse on $\Sigma_\pm$ are
\ba
&& \gamma_{AB}=g_{ab} Y^a{}_{,A} Y^b{}_{,B}\, ,\\
&&d\gamma^2=\gamma_{AB} \dd z^A \dd z^B=\ts{\xi}^2 \dd v^2-2 \dd v \dd\lambda_{\pm}\, ,\\
&& \partial_\gamma^2 = \gamma{}^{AB} \partial_{z{}^A} \partial_{z{}^B} = -2 \partial_v \partial_{\lambda_{\pm}} - \ts{\xi}^2 \partial_{\lambda_{\pm}}^2 \, .
\ea
One also has $\sqrt{-\gamma} = 1$.

Let us denote by $n{}^\mu_{(i)}$ the $D-2$ mutually orthogonal unit vectors  normal to $\Sigma_{\pm}$.
We define the extrinsic curvature $\Omega{}_{(i)AB}$ by the relations
\be
\Omega{}_{(i)AB}= g{}_{ab} n{}_{(i)}^a Y{}^c{}_{,A} \nabla{}_c Y{}^b{}_{,B} \, .
\ee
The surface is called minimal if the trace of the extrinsic curvature
\be
\Omega{}_{(i)} \equiv \gamma{}^{AB}\Omega{}_{(i)AB}
\ee
vanishes. This condition can be written in the form
\ba
&&\Omega{}_{(i)}=g_{ab} n_{(i)}^a Z^b=0\, ,\\
&&Z^b=\gamma{}^{AB} Y{}^c{}_{,A} \nabla{}_c Y{}^b{}_{,B}\, .\n{ZZZ}
\ea
Thus if the vector $\ts{Z}$ is orthogonal to all external vectors $\ts{n}_{(i)}$ the surface $\Sigma_{\pm}$  is minimal, and hence it is a solution of the Nambu--Goto equation and defines a worldsheet of a stationary string. This happens when $\ts{Z}$ is tangent to $\Sigma_{\pm}$. Let us show that this condition is indeed satisfied.
Using (\ref{YYY}) and (\ref{ZZZ}) one gets
\be
Z{}^b = -\left( \xi{}^a\nabla{}_a \ell{}_{\pm}^b + \ell{}_{\pm}^a \nabla{}_a \xi{}^b + \ts{\xi}^2\ell{}_{\pm}^a \nabla{}_a \ell{}_{\pm}^b\right)\, .
\ee
Since $\ts{\ell}_{\pm}$ is a tangent vector to affinely parametrized null geodesic, the last term in the parentheses vanishes. Equation (\ref{xll}) implies that the second term in the parentheses is equal to the first one. Thus
\be
Z^b=2\ell_{\pm}^a \nabla_a \xi^b=-F^b_{\ \ a}\ell_{\pm}^a=-\kappa_{\pm}\ell_{\pm}^b\, .
\ee
The last relation demonstrates that $\ts{Z}$ is tangent to $\Sigma_{\pm}$ and hence the latter is a minimal surface.

\subsection{Principal Killing strings in Kerr--NUT--(A)dS spacetime}
\label{sec:off-shell}

Let us emphasize that a principal Killing string is uniquely defined by the principal tensor. Since Kerr--NUT--(A)dS spacetimes admit such a tensor, our construction automatically produces special stationary solutions of the Nambu--Goto equations in the Kerr--NUT--(A)dS metrics in any number of dimensions. In this section we describe these solutions.

We denote
\ba
&&\ts{e}^n={1\over \sqrt{2}}(\ts{l}_+ -\ts{l}_-)\hh
\hat{\ts{e}}^n={1\over \sqrt{2}}(\ts{l}_+ +\ts{l}_-)\, ,\\
&& (\ts{e}^n)^2=1\hhh (\hat{\ts{e}}^n)^2=-1\hhh (\ts{e}^n,\hat{\ts{e}}^n)=0\, .
\ea
Then (\ref{gg}) takes the form
\be\n{mee}
\ts{g}=-\ts{e}^{n}\ts{e}^{n}+ \hat{\ts{e}}^{n}\hat{\ts{e}}^{n} +\sum\limits_{\mu=1}^{n-1} (\ts{e}^{\mu}\ts{e}^{\mu}+ \hat{\ts{e}}^{\mu}\hat{\ts{e}}^{\mu})+\epsilon \hat{\ts{e}}^{0}\hat{\ts{e}}^{0}\, .
\ee

For the most general Lorentzian metric admitting a (non-degenerate) principal  tensor $\ts{h}$ in $D = 2n + \epsilon$ dimensions one has \cite{Chen:2006xh,Kubiznak:2008zs,Frolov:2017kze}
\ba
&&\ts{e}^n={\dd r\over \sqrt{Q_n}}\hh \hat{\ts{e}}^n=\sqrt{Q_n}\sum_{j=0}^{n-1} A_n^{(j)} \dd\psi_j\, ,\nonumber\\
&&\ts{e}^{\mu}={\dd x_{\mu}\over \sqrt{Q_{\mu}}}\hh \hat{\ts{e}}^{\mu}=\sqrt{Q_{\mu}}\sum_{j=0}^{n-1}A_{\mu}^{(j)} \dd\psi_j\, ,\nonumber\\
&&\hat{\ts{e}}^0=\sqrt{Q_0}\sum_{j=0}^{n} A^{(j)} \dd\psi_j\hh
Q_0=-{c\over A^{(n)}}\, .
\ea
In the above we defined
\ba
&&A^{(j)}_\mu = \sum\limits_{\substack{\nu_1,\dots,\nu_k=1 \\ \nu_1 < \dots < \nu_j \\\nu_i \not= \mu }}^n x{}_{\nu_1}^2 \dots \, x{}_{\nu_j}^2\, ,\\
&&A^{(j)}=\sum_{\nu_1<\ldots<\nu_j} x_{\nu_1}^2\ldots x_{\nu_j}^2\hhh Q_{\mu}={X_{\mu}\over U_{\mu}}\, ,\\
&& U_{\mu}=-(r^2+x_{\mu}^2)\prod_{\substack{\nu=1 \\ \nu\neq \mu}}^{n-1} (x_{\nu}^2-x_{\mu}^2)\, ,\\
&& U_{n}=\prod_{\nu=1}^{n-1} (r^2+x_{\nu}^2)\hhh x_n^2=-r^2\, .
\ea
Here the range of Greek indices is $\mu,\nu=1,\ldots,n-1$ while Latin indices take values $k,l=0,\ldots,n-1+\epsilon$. $\tau\equiv \psi_0$ is the time coordinate.  The constant $c$ appearing in odd dimensions is an undetermined parameter.

The metric (\ref{mee}) is called \emph{off-shell} if the functions $X{}_\mu$ are left unspecified. Invoking the Einstein equations with cosmological constant yields a suitable choice for the $X_\mu$ that reproduces the Kerr--NUT--(A)dS class of spacetimes for any $D = 2n + \epsilon$ \cite{Frolov:2017kze}.

In the metric \eqref{mee} the principal null vectors $\ts{\ell}_\pm$  take the form \cite{Kubiznak:2008zs} ($m \equiv n - 1 + \epsilon$)
\begin{align}
\label{eq:principal-null-congruence}
\ts{\ell_\pm} = \pm \partial_r + \frac{r^{2(n-1)}}{X_n} \partial_\tau + \frac{1}{X_n} \sum\limits_{j=1}^m r{}^{2(n-1-j)}\partial{}_{\psi_j} \, .
\end{align}
Both null vectors $\ts{\ell}_{\pm}$ are future-directed. The radial coordinate $r$ increases along $\ts{\ell}_+$ and decreases along $\ts{\ell}_-$. We call a null geodesic generated by $\ts{\ell}_+$ an out-going null ray, while  similar curves for $\ts{\ell}_-$ are in-coming null rays. By comparing \eqref{eq:principal-null-congruence} with \eqref{lrr} one can conclude that for $\ts{\ell}_{\pm}$ one has $\lambda_{\pm}=\pm r$. In the presence of a black hole, in-coming null rays enter the future event horizon $\mathcal{H}_+$ and remain regular at it. Similarly, the out-going null rays are regular at the past event horizon $\mathcal{H}_-$.

In what follows we consider a case when  a spatially infinite string crosses the black hole horizon.  A corresponding solution should be regular at $\mathcal{H}_+$, and we assume that after some relaxation processes the string becomes stationary. In such a configuration there exist two antipodal pieces (segments) of the string piercing the black hole. We assume that each of them is described by the principal Killing surface, regular at $\mathcal{H}_+$, that is, by $\Sigma_-$. For this reason, from now on we always assume that our string-segment solution is generated by vectors $\ts{\xi}$ and $\ts{\ell}_-$ and omit the subscript ``$-$'' in the corresponding notations.

The principal null ray equations
\be
{dx^a\over d\lambda}=\ell^a
\ee
can be easily solved, yielding $\lambda=-r$. Let us denote
\be
P_n^{(j)}=\int {r^{2(n-1-j)} dr\over X_n(r)}\, .
\ee
Then, in $\{\tau,r,x_{\mu},\psi_j\}$ coordinates, one finds
\be
\label{eq:pks-main-result}
\tau=-P_n^{(0)}(r)\hhh \psi_j=-P_n^{(j)}(r)\hhh x_{\mu}= \text{const} \, .
\ee

Let us perform a following coordinate transformation $\{\tau,r,x_{\mu},\psi{}_k\} \rightarrow \{v,r, x_{\mu},\hat{\phi}_k\}$ with $k=1,\dots,m=n-1+\epsilon$ according to
\begin{align}
\begin{split}\n{cct}
\dd \tau &= \dd v  - \frac{r^{2(n-1)}}{X_n} \, \dd r \, , \\
\dd \psi_j &= \, \dd \hat{\phi}_j - \frac{1}{X_n} r^{2(n-1-j)} \, \dd r \, ,
\end{split}
\end{align}
In these coordinates the principal Killing string equation takes the form
\be
\hat{\phi}_j=\hat{\phi}_j^{0}=\text{const}\hh x_{\mu}=x^0_{\mu}=\text{const}\, ,
\ee
while $v$ and $r$ are arbitrary. In other words, the transformation (\ref{cct}) ``straightens'' the string.

\subsection{Stress-energy tensor of the principal Killing string}

The dynamics of a test string in an external gravitational field $g_{ab}$ is described by the Nambu--Goto action
\be
I=-\mu_s \int d^2\zeta \sqrt{- \det\left(\gamma_{AB}\right)}\, ,\ \
\gamma_{AB}= g_{ab}Y^{a}{}_{,A} Y^{b}{}_{,B}\, .
\ee
$\mu_s$ is the string tension and $\zeta^A$ $ (A=0,1)$ are coordinates on the string world-sheet. The functions $Y^{a}(\zeta^A)$ determine the string's embedding in the bulk spacetime. The stress-energy tensor of the string is localized on its surface and is of the form \cite{vilenkin2000cosmic}
\be\n{Tmn}
T^{ab}= {-\mu_s\over \sqrt{-g}} \int d^2\zeta \sqrt{-\gamma} \gamma^{AB} Y^{a}{}_{,A} Y^b{}_{,B} \delta^{(4)}(x^{c}-Y^{c}(\zeta^B))\,  .
\ee
In $\{v,r,x_{\mu},\hat{\phi}_j\}$ coordinates this tensor reads
\begin{align}
T{}^{ab} = \frac{\mu_s}{\sqrt{-g}} \left( 2\xi{}^{(a} \ell{}^{b)} + \ts{\xi}^2 \ell{}^a \ell{}^b \right ) q \, , \label{STT}
\end{align}
were we defined
\begin{align}\n{QQQ}
q = q(x_\mu ,\hat{\phi}_j | x^0_\mu, \hat{\phi}^0_j) \equiv \prod\limits_{\mu=1}^{n-1} \delta(x_\mu - x_\mu^0) \prod\limits_{j=1}^m \delta(\hat{\phi}_j - \hat{\phi}_j^0) .
\end{align}
The expressions (\ref{STT}) and (\ref{QQQ}) allow one to express the stress-energy tensor of a principal Killing string purely in terms of geometrical quantities in the off-shell Kerr--NUT--(A)dS metric \eqref{mee}. For applications, the on-shell version of these relations is more useful. Certainly, the above formulas work for this case as well.
In particular, one can use the on-shell version of (\ref{STT}) for the calculations of the energy and angular momentum fluxes through the horizon of the Kerr--NUT--(A)dS black hole. However, such calculations might be quite involved. For the off-shell metrics one must first formulate conditions when the horizon is regular.

The other problem is that the Killing coordinates $\psi_i$ are connected to the secondary Killing vectors determined by the principal tensor. In general, these coordinates differ from the Boyer--Lindquist coordinates $\phi_i$, and are related to the latter by means of linear transformations.\footnote{For the discussion of these relations, see \cite{Frolov:2017kze}.} In particular, the Killing vectors $\partial_{\phi_i}$ associated with these coordinates $\phi_i$ enter the standard definition of the angular momentum.
Finally, when and if the energy and angular momentum fluxes through the horizon are properly defined and calculated, one would need to define how they are connected with the change of the parameters of the Kerr--NUT--(A)dS solution, when the backreaction effects are taken into account. Therefore, in this work, we shall restrict ourselves to illustrating the properties of principal Killing strings in Myers--Perry spacetimes.

\section{Principal Killing strings in Myers--Perry spacetimes}
\label{sec:myers-perry}

\subsection{Myers--Perry black holes}
In this section we illustrate the obtained results for the special case of the $D$-dimensional Myers--Perry black hole vacuum solution \cite{Myers:1986un,Myers:2011yc}. By using the Komar definition of the mass and angular momentum we reproduce the results of \cite{Frolov:2004qw} for the angular momentum transfer from a string piercing the black hole. We relate this effect to the action of a torque produced by a stationary string at spatial infinity.

Let us begin by briefly reviewing some general properties of the Myers--Perry metric in $D = 2n + \epsilon$ spacetime dimensions. It generalizes the Kerr solution to higher dimensions, and as such can be expressed in terms of the canonical coordinates $\{\tau, r, x{}_\mu,\psi_i \}$, see e.g.\ Ch.~4 in Ref.~\cite{Frolov:2017kze}. For our purposes, however, it is more convenient to express it in terms of the Myers--Perry coordinates $\{t, r, \mu_i, \phi_i\}$ (in units where $c=1$ and $G=1$):
\begin{align}
\ts{g} &= -\dd t^2 + \frac{U \, \dd r^2}{V - 2\mathcal{M}} + \frac{2\mathcal{M}}{U}\bigg( \dd t - \sum\limits_{i=1}^m a_i \mu_i^2 \dd \phi_i \bigg)^2 \label{eq:myers--perry-metric} \\
&\hspace{12pt} + \sum\limits_{i=1}^m \left(r^2+a_i^2\right)\left(\dd\mu_i^2 + \mu_i^2 \dd \phi_i^2\right) + (1-\epsilon) r^2 \dd \mu_0^2 \, , \nonumber
\end{align}
where we defined the auxiliary functions
\begin{align}
\begin{split}
V &= \frac{1}{r^{1+\epsilon}} \prod\limits_{i=1}^m \left(r^2 + a_i^2\right)\, , \\
U &= V\bigg(1 - \sum\limits_{i=1}^m \frac{a_i^2\mu_i^2}{r^2 + a_i^2} \bigg) \, .
\end{split}
\end{align}
As earlier, we denote $m \equiv n - 1 + \epsilon$, and the number of spatial dimensions is $D-1=2m+(1-\epsilon)$. Note that despite of its appearance the metric \eqref{eq:myers--perry-metric} is not diagonal in the coordinates $\mu_i$ because these ``directional cosines'' are constrained via
\begin{align}\n{constr}
\sum\limits_{i=\epsilon}^m \mu_i^2 = 1 \, .
\end{align}
The spatial coordinates in the metric (\ref{eq:myers--perry-metric}) are $r$, $\mu_i$, $\phi_i$ and (in the case $\epsilon=0$) one more coordinate $\mu_0$. Their total number is $2m+1+(1-\epsilon)$. This counting correctly reproduces the required number of spatial dimensions if one takes into account the constraint (\ref{constr}).

The Myers--Perry solution contains  the mass parameter $\mathcal{M}$ as well as $m$ angular momentum parameters $a_i$. Its isometries are encoded by the timelike Killing vector $\ts{\xi} = \partial_t$ as well as $m$ mutually commuting azimuthal Killing vectors $\ts{\zeta{}_{(i)}} = \partial_{\phi_i}$. The ingoing principal null congruence \eqref{eq:principal-null-congruence} then takes the form
\begin{align}
\ts{\ell} = - \partial_r + \frac{V}{V - 2\mathcal{M}} \bigg( \partial_t  + \sum\limits_{i=1}^m \frac{a_i}{r^2 + a_i^2} \partial_{\phi_i} \bigg) \, . \label{eq:myers-perry-pnc}
\end{align}
The horizon of the Myers--Perry black hole is located at $V(r_h) = 2\mathcal{M}$ which corresponds to
\begin{align}
\prod\limits_{i=1}^m (r_h^2 + a_i^2) = 2\mathcal{M} r_h^{1+\epsilon} .
\end{align}
Since the Myers--Perry coordinates are not regular at the horizon, it is useful to instead work with the coordinates $\{v, \hat{\phi}_i\}$ related to $\{t, \phi_i\}$ via
\begin{align}
\dd v &= \dd t + \frac{V}{V-2\mathcal{M}} \dd r \, , \\
\dd \hat{\phi}_i &= \dd \phi_i + \frac{V}{V-2\mathcal{M}} \frac{a_i}{r^2 + a_i^2} \dd r \, .\n{fff}
\end{align}
The Myers--Perry metric then takes the form
\begin{align}
\label{eq:mp-regular-ingoing}
\ts{g} &= - \dd v^2 + 2 \dd v \dd r - 2 \sum\limits_{i=1}^m a_i \mu_i^2 \dd \hat{\phi}_i \dd r \nonumber \\
&\hspace{12pt} + \frac{2\mathcal{M}}{U} \bigg( \dd v - \sum\limits_{i=1}^m a_i \mu_i^2 \dd \hat{\phi}_i \bigg)^2  \\
&\hspace{12pt}+ \sum\limits_{i=1}^m \left( r^2 + a_i^2 \right)\left( \dd \mu_i^2 + \mu_i^2 \dd \hat{\phi}_i^2 \right) + (1 - \epsilon)r^2 \dd \mu_0^2 \, , \nonumber
\end{align}
which is manifestly regular at the horizon. The principal null congruence and the Killing vectors are
\begin{align}
\label{eq:mp-pnc-killing-vectors}
\ts{\ell} = -\partial_r \, , \quad \ts{\xi} = \partial_v \, , \quad \ts{\zeta{}_{(i)}} = \partial{}_{\hat{\phi}_i} \, .
\end{align}
In the coordinates $\{ v,r,\mu_i,\hat{\phi}_i \}$, the equation of the principal Killing string takes the simple form $\{\mu_i,\hat{\phi}_i\}= \text{const}$. Integrating Eq.~\eqref{fff} for $\hat{\phi}_i=\text{const}$ one obtains the following relations that determine the form of the principal Killing string in the Myers--Perry coordinates:
\be\n{FiFi}
\phi_i=\phi^0_i-a_i\int {V dr\over (V-2\mathcal{M}) (r^2+a_i^2)}\, .
\ee
The integrals in these relations are logarithmically divergent at the event horizon. This means that the string makes an infinite number of revolutions around the black hole. Let us emphasize that this is a pure kinematic effect connected with the time delay near the horizon. In $\{v,r,\mu_i,\hat{\phi}_i\}$ coordinates, which are regular at the horizon, the string is straightened and enters the horizon without any turns.

\subsection{Mass and angular momentum of the Myers-Perry black hole}

The mass $M$ and the angular momenta $J_{(i)}$ of the Myers--Perry black hole are given by the following Komar integrals \cite{Myers:1986un}:
\begin{align}
\begin{split}
-16 \pi \frac{D-3}{D-2} M &= \oint\limits_{\mathcal{B}}
 \nabla{}^a \xi{}^b \dd S{}_{ab} \, , \label{eq:horizon-mass-angular-momentum} \\
-16 \pi J_{(i)} &= \oint\limits_{\mathcal{B}}  \nabla{}^a \zeta{}_{(i)}^b \dd S{}_{ab} \, .
\end{split}
\end{align}
Here we follow the conventions of Ref.~\cite{Poisson:2004}, Ch.~3. Here $\mathcal{B}$
is a spacelike $(D-2)$-dimensional section of the event horizon, and $\dd S{}_{ab}$ is a surface element of $\mathcal{B}$. $\ts{\xi}$ is the timelike Killing vector, and $\ts{\zeta{}_{(i)}}$ denotes the rotational Killing vectors related to the axisymmetry.

For an isolated stationary vacuum black hole these parameters do not depend on the choice of the section $\mathcal{B}$. In what follows we denote by $\mathcal{B}_v$  the intersection of the null surface $v= \text{const}$ and the horizon surface. The corresponding quantities represent the mass and components of the angular momentum at the moment of advanced time $v$.

The mass $M$ and angular momenta $J_{(i)}$ are related to the parameters $\mathcal{M}$ and $a_i$ via \cite{Myers:2011yc}
\begin{align}
\begin{split}
M &= \frac{(D-2)\omega_{D-2}}{8\pi} \mathcal{M} \, \\
J_{(i)} &= \frac{\omega_{D-2}}{4\pi}\mathcal{M} a_i = \frac{2}{D-2} M a_i \, ,
\end{split}
\end{align}
where $\omega_{D-2} = 2 \pi^{(D-1)/2}/\Gamma((D-1)/2)$ is the surface of the $S^{D-2}$ sphere. The normalization is chosen such that for $D=4$ one recovers the usual relations $\mathcal{M} = M$ and $J = M a$.

The Komar definition of the mass and angular momentum, as measured at infinity, implies
\begin{align}
\begin{split}\n{KOMAR}
-16\pi\frac{D-3}{D-2} M^\infty_{t} &= \oint\limits_{S^{D-2}_{\infty,t}}
\!\!\! \nabla{}^a \xi{}^b \dd S{}_{ab} \, , \\
-16\pi J^\infty_{(i) t} &= \oint\limits_{S^{D-2}_{\infty,t}} \!\!\! \nabla{}^a \zeta{}_{(i)}^b \dd S{}_{ab} \, .
\end{split}
\end{align}
Here the integration is performed at the given time $t$ over $(D-2)$-dimensional sphere of large radius $r$ surrounding the black hole, and $\dd S{}_{ab}$ is a surface element on this sphere.

If the black hole exterior is empty then the mass and angular momentum measured at infinity, (\ref{KOMAR}), coincide with (\ref{eq:horizon-mass-angular-momentum}) \cite{Myers:1986un}. In order to prove this it is sufficient to use the following integrability condition which is valid for any Killing vector $K$,
\begin{align}
\nabla{}^a \nabla{}_a K{}^b = -R{}^b{}_a K{}^a \, . \label{eq:integrability-killing-vector}
\end{align}
By integrating this equality with $R{}^b{}_a=0$ over a $(D-1)$-dimensional surface crossing the horizon and extended to infinity and using Stokes' theorem one can show that expressions \eqref{eq:horizon-mass-angular-momentum} and \eqref{KOMAR} give the same value for the mass and angular momentum. In the presence of a cosmic string the mass at infinity differs from the black hole mass, the difference being the contribution of the string's mass $M_s$. For a straight string of length $L$ the mass is $M_s=\mu_s L$, so that for an infinite string this mass becomes infinitely large. In our consideration we use a test string approximation which implies that $M_s$ is much smaller than the mass of the black hole. In this approximation we assume that the string has finite size. For example, it may have an end point equipped with a monopole and an external force is applied to the latter in order to keep the string in equilibrium. For a finite length $L$ of the string its mass $M_s$ can be made arbitrary small by proper choice of the string's tension $\mu_s$. In what follows we assume that the required test string approximation is valid. In fact, we shall study only the change of the black hole parameters with time, and the quantity $M_s$ does not enter the results.

\subsection{Energy and angular momentum fluxes}
Let us introduce the new coordinate
\be
T=v-r\, ,
\ee
and consider a $(D-1)$-dimensional surface $\Sigma$ where $T=\text{const}$. This surface is spacelike both outside and at the horizon. On the horizon this new time $T$ differs from the advanced null time $v$ only by an additive constant $r_h$.

In $D$ spacetime dimensions, the functions $V(r)$ and $U(r)$, which enter the metric, have the following asymptotics at $r \rightarrow \infty$:
\begin{align}
U(r)\sim V(r) \sim r^{2m-1-\epsilon} = r^{D - 3}  \,  .
\end{align}
In what follows we assume that $D\ge 5$. Since
\be
\dd T=\dd t + {2\mathcal{M}\over V-2\mathcal{M}} \dd r\, ,
\ee
the coordinate $T$ coincides with the asymptotic Killing time $t$ at infinity up to a constant.

We denote by $\Sigma_{T_0}$ a $(D-1)$-dimensional surface which is determined by the equation $T=T_0=\text{const}$. Moreover, we assume that the time $T_0$ is chosen such that a possible time dependent evolution of the string has finished before this time. In other words, in the spacetime domain where $T>T_0$, the string is stationary and coincides with the principal Killing solution, which was described earlier\footnote{Strictly speaking, and in the absence of any external forces, if the test string is exactly stationary in this time interval it would be stationary forever. If a dynamical string pierces the black hole at some earlier time, then just after this happens there would be string excitations propagating along the string in both directions: to the horizon and to infinity. Our  assumption is that after $T_0$ one can neglect this relaxation processes and use a stationary string approximation.}. We shall call a one-dimensional line obtained by the intersection of the string with the surface $T=\text{const}$ a string configuration at time $T$.

As we shall demonstrate later, in the presence of a string the parameters of the black hole, and hence its gravitational field, will slightly change in time. Let us neglect this effect for the moment and consider two surfaces $\Sigma_1$ and $\Sigma_2$, defined by the equations $T=T_1=v_1-r_h$ and $T=T_2=v_2-r_h$, respectively, assuming that $v_2>v_1>T_0+r_h$. In other words, we chose the time interval $T_2-T_1$ to be much smaller than the characteristic time of the black hole evolution.
The surfaces $\Sigma_1$ and $\Sigma_2$ intersect the horizon at two $(D-2)$-dimensional surfaces ${\cal B}_1$ and ${\cal B}_2$. Denote by ${\cal H}_{1,2}$ a part of the event horizon between ${\cal B}_1$ and ${\cal B}_2$, and by ${\cal H}_0$ a surface determined by the equation $r=r_0>r_h$ that lies between $\Sigma_1$ and $\Sigma_2$. Finally, we denote by $V$ the $D$-dimensional volume restricted by the four surfaces $\Sigma_1$, $\Sigma_2$, ${\cal H}_{1,2}$ and ${\cal H}_0$.

The background Myers--Perry metric has symmetries generated by the Killing vectors $\ts{\xi}$ and $\ts{\zeta}_{(i)}$. For a given stress-energy tensor $T_{ab}$ one can define conserved Killing currents
\be\n{jjjj}
j_{\xi}^a=T^{ab}\xi_b\, ,\
j_{\zeta_{(i)}}^a=T^{ab}\zeta_{(i) b}\, , \
j_\xi^a{}_{;a}=j_{\zeta_{(i)}}^a{}_{;a}=0\, .
\ee

Since the Killing currents are conserved, the integrals of $j_{\xi\ ;a}^a$ and $j_{\zeta_{(i)}}^a{}_{;a}$ over the $D$-volume $V$ vanish identically. Stokes' theorem implies that
\begin{align}
\int\limits_V j^{a}{}_{;a} \sqrt{-g} \, \dd^D x=\int\limits_{\partial V} j^a \dd\hat{\Sigma}_a\, .
\end{align}
Denote by $x^a$ the coordinates in the bulk space $V$ and by $y^p$ the coordinates on its  boundary $\partial V$. Then $x^a=x^a(y^p)$ is the defining equation of $\partial V$. The volume element $\dd\hat{\Sigma}_a$ is defined as follows (see e.g.~\cite{frolov1998black}):
\begin{align}
\dd\hat{\Sigma}_a={1\over (D-1)!} e_{a b_1\ldots b_{D-1}} \det\left( {\partial x^{b_i}\over \partial y^{p_j}}\right) \, \dd^{D-1}y\, ,
\end{align}
where $e_{a b_1 \dots b_{D-1}}$ denotes the completely antisymmetric tensor. This definition is valid for any smooth boundary, provided the orientation of the coordinates in $V$ and at $\partial V$ are chosen properly (for the discussion of this point see e.g.~\cite{Poisson:2004}).
In the above definition, in case of spacelike and timelike surfaces $\partial V$, the surface element $\dd\hat{\Sigma}_a$ is directed towards the exterior of $V$. If a piece of the boundary is null, the definition of $\dd\hat{\Sigma}_a$ requires additional specification, see \cite{Poisson:2004} for a detailed discussion. We shall provide the corresponding explicit expression for $\dd\hat{\Sigma}_a$ for a piece of the null boundary at the horizon in the next subsection.

In what follows, it is convenient to chose surface elements on $\Sigma_1$ and $\Sigma_2$ to be both future-directed. The corresponding surface element $\dd\Sigma_a$ on $\Sigma_2$ coincides with $\dd\hat{\Sigma}_a$, while on $\Sigma_1$ it has the opposite sign. We also choose $\dd\Sigma_a$ on the external boundary ${\cal H}_0$ to be directed inwards, and hence it also differs by sign from $\dd\hat{\Sigma}_a$. Using these definitions and Stokes' theorem for the conserved currents one obtains the following relations:
\ba
&& \!\!\!\! \Big[\int_{ {\cal H}_{1,2}} \!\!-\! \int_{ {\cal H}_0}\Big] j_{\xi}^a \dd \Sigma_a=
\Big[\int_{ \Sigma_{1}}\!\!-\!\int_{\Sigma_{2}} \!\Big] \ j_{\xi}^a \dd\Sigma_a \, ,\n{HSxi}\\
&& \!\!\!\! \Big[\int_{ {\cal H}_{1,2}} \!\!-\! \int_{ {\cal H}_0}\Big] j_{\zeta_{(i)}}^a \dd\Sigma_a=
\Big[\int_{ \Sigma_{1}} \!\!-\! \int_{\Sigma_{2}} \! \Big] j_{\zeta_{(i)}}^a \dd\Sigma_a\, .\n{HSze}
\ea

Since the surface $\Sigma_2$ is obtained by a rigid shift of $\Sigma_1$, these two surfaces are isometric and for a stationary string the expressions in the right-hand sides of Eqs.~\eqref{HSxi} and \eqref{HSze} vanish. As a result one has
\begin{align}
\int_{ {\cal H}_{1,2}} j_{\xi}^a \dd\Sigma_a &= \int_{ {\cal H}_0}j_{\xi}^a \dd\Sigma_a\, ,\\
\int_{ {\cal H}_{1,2}} j_{\zeta_{(i)}}^a \dd\Sigma_a &= \int_{ {\cal H}_0} j_{\zeta_{(i)}}^a \dd\Sigma_a\, .
\end{align}
Thus, the energy and momentum fluxes through the part ${\cal H}_{1,2}$ of the horizon, generated by the string, are equal to the similar fluxes through the external boundary ${\cal H}_0$. 

\subsection{Energy and momentum fluxes through the horizon}

We demonstrate now that in the presence of the string that enters the horizon there is angular momentum transfer into the black hole. Let us first calculate the corresponding fluxes of the energy and angular momentum in the adopted test field approximation and discuss the backreaction of these fluxes on the black hole geometry later.

The energy and angular momentum fluxes through the horizon during the time interval $(v_1,v_2)$ are
\begin{align}
\Delta E&=\int_{{\cal H}_{1,2}} j_{\xi}^a \dd\Sigma_a\, ,\n{EEE}\\
\Delta J_{(i)}&=\int_{{\cal H}_{1,2}} j_{\zeta_{(i)}}^a \dd\Sigma_a\,.\n{JJJ}
\end{align}
We shall now apply these relations to the stress-energy of the test string, Eqs.~\eqref{STT}--\eqref{QQQ}, adapted to the Myers--Perry metric.

First of all, we assume that the constraint \eqref{constr} is resolved and write $\mu_i=\mu_i(\omega_k)$, where $k=\epsilon,\ldots,m-1$. In what follows, the explicit form of new unconstrained coordinates $\omega_k$ is not important. Moreover, for our calculations we adopt the incoming null coordinates $\{v,r,\omega_k,\hat{\phi}_i\}$.
In these coordinates, the future-directed volume element $\dd \Sigma_a$ on the horizon $\mathcal{H}_{1,2}$ can be obtained as follows: Following \cite{Poisson:2004}, we define the function $\Phi$ that increases towards the future and vanishes on the horizon,
\begin{align}
\Phi \equiv r_h - r \, .
\end{align}
Then, the following vector is  null and future-directed on the horizon
\begin{align}
k_a \equiv -\partial_a \Phi = r_{,a} = \delta{}^r_a \ .
\end{align}
Equation \eqref{eq:mp-pnc-killing-vectors} shows that the future directed  principal null vector $\ts{\ell}$ satisfies the condition
\begin{align}
(\ts{k}, \ts{\ell} ) = -1 \, .
\end{align}
Using these two vectors one can write the inverse metric on the horizon as follows:
\be
g^{ab} = -k^a \ell^b-k^b \ell^a + \sigma^{ab}\, ,
\ee
where $\sigma^{ab}$ is the inverse of the $(D-2)$-dimensional metric induced on a spacelike slice $S$ of the horizon which is orthogonal to both $\ts{k}$ and $\ts{\ell}$. One also has on the horizon $\sqrt{-g}=\sqrt{\sigma}$, where $\ts{\sigma}$ is the metric on $S$. Let us denote
\be
\dd^{D-2}\Omega=\prod\limits_{k=\epsilon}^{m-1} \dd \omega_k \prod\limits_{i=1}^m \dd \hat{\phi}_i\, .
\ee
Then, according to \cite{Poisson:2004}, a properly oriented surface element on the horizon is
\begin{align}
\begin{split}
\label{eq:volume-surface-elements}
\dd \Sigma{}_a &= k^b 2 k_{[a} \ell_{b]} \sqrt{-g} \, \dd v \, \dd\Omega^{D-2} \\
&= - \sqrt{-g} r_{,a} \, \dd v \, \dd\Omega^{D-2}=-k_a \sqrt{\sigma} \, \dd v \, \dd\Omega^{D-2}\, .
\end{split}
\end{align}

The stress-energy of a principal Killing string in $\{v,r,\omega_k,\hat{\phi}_i\}$ coordinates is
\be
T{}^{ab} = \mu_s \left( 2\xi{}^{(a} \ell{}^{b)} + \ts{\xi}^2 \ell{}^a \ell{}^b \right ) \, q \, , \label{eq:stress-energy}
\ee
where $q$ is given by
\be\n{qqq}
q={1\over \sqrt{-g}} \prod\limits_{k=\epsilon}^{m-1} \delta( \omega{}_k -\omega{}_k^0) \prod\limits_{j=1}^m \delta(\hat{\phi}_i-\hat{\phi}_i^0)\, .
\ee
Here $\omega{}_k^0$ and $\hat{\phi}_i^0$ are the angular coordinates of the string.

We can now perform the integrals (\ref{EEE})--(\ref{JJJ}) over the part of the horizon between its slices $\mathcal{B}_1$ and $\mathcal{B}_2$ to obtain
\begin{align}
\Delta E &= 0\hh
\Delta J_{(i)} =\dot{J}_{(i)}(v_2 - v_1)\, ,\n{EEJJ}\\
\dot{J}_{(i)}&= -\mu_s a_i \left( \mu_i^0 \right)^2 \, .\n{dJJ}
\end{align}
In other words, there is no energy (mass) flux from the string to the black hole, while the angular momentum transfer does not vanish and its rate $\dot{J}_{(i)}$ is permanent in time. These relations correctly reproduce the results of Ref.~\cite{Frolov:2004qw}.

\subsection{Spatial infinity}

Since the energy and angular momentum fluxes through the external boundary ${\cal H}_0$ do not depend on the choice of its radius $r_0$ one can make this radius arbitrary large.
In this domain, $\Sigma_1$ and $\Sigma_2$ coincide with the surfaces of constant time $t$. It is instructive to repeat the calculations of the energy and momentum fluxes at infinity, that is, in the $r_0\to\infty$ limit.

The asymptotic form of the Myers--Perry metric \eqref{eq:myers--perry-metric} is
\begin{align}
\begin{split}
\ts{g} &= -\dd t^2 + \dd r^2 + \sum\limits_{i=1}^m r^2 \left(\dd\mu_i^2 + \mu_i^2 \dd \phi_i^2\right) \\
&\hspace{12pt} + (1-\epsilon) r^2 \dd \mu_0^2 \, , \label{eq:minkowski}
\end{split}
\end{align}
which corresponds to the $D$-dimensional Minkowski metric in multi-polar coordinates \cite{Myers:2011yc}. The null congruence $\ts{\ell}$ simplifies to
\begin{align}
\ts{\ell} = - \partial_r + \partial_t  +{1\over r^2} \sum\limits_{i=1}^m a_i \partial_{\phi_i} \, . \label{eq:pnc-mp-asymptotic}
\end{align}
Using relation \eqref{FiFi}, one finds that for large $r$ the principal Killing string configuration has the form
\be\n{phiass}
\phi_i=\phi_i^0 +{a_i\over r}+\ldots \, .
\ee
The future-directed, radially inward-pointing surface element of ${\cal H}_{\infty}$ at $r_0=\infty$ is
\be
\dd \Sigma{}^\infty_a = -\sqrt{-g} \, \delta{}^r_a \, \dd t \, \dd ^{D-2}\Omega\, ,
\ee
where $\sqrt{-g} = r^{D-2} \sigma^{D-2}$, and $\sigma^{D-2}$ denotes the angular part of the determinant. The stress energy tensor of the string is \eqref{eq:stress-energy} with the asymptotic form of $\ts{\ell}$ as given in Eq.~\eqref{eq:pnc-mp-asymptotic} such that
\begin{align}
T{}^{ab} = \mu_s \left( \delta{}^a_t \delta{}^b_t - \delta{}^a_r \delta{}^b_r - \sum\limits_{i=1}^m \frac{2 a_i}{r^2} \delta{}^{(a}_r \delta{}^{b)}_{\phi_i} +  \dots \right) \, q \, .
\end{align}
Here $q$ is the same as (\ref{qqq}) with the only changes $\hat{\phi}_i\to \phi_i$ and $\hat{\phi}_i^0\to \phi_i^0$. Calculations give
\begin{align}
\begin{split}
&\Delta E^\infty =0\hh \Delta J^\infty_{(i)}=\dot{J}^\infty_{(i)}(t_2-t_1)\, ,\\
&\dot{J}^\infty_{(i)}=-\mu_s a_i \left( \mu_i^0 \right)^2\, .\n{MJINF}
\end{split}
\end{align}
As expected, this result is consistent with the calculation for the mass and angular momentum fluxes through the horizon, corresponding to an extraction of angular momentum.

\subsection{Backreaction}
Let us now give a few comments concerning the backreaction of the fluxes through the horizon generated by the string. The ``common sense'' expectation is that---as a result of these fluxes---the parameters of the black hole should change. However, the demonstration of this requires the development of the corresponding formalism. The main problem is the following: In vacuum, in the absence of fluxes, the event horizon coincides with the Killing horizon, which can be defined at any moment of time. In the presence of fluxes, however, the metric becomes time-dependent, and strictly speaking there is no required Killing vector which allows one to define the black hole boundary at a given moment. Because of its teleological nature, the standard definition of the event horizon is also not very useful for this purpose.

Several generalizations to situations when the parameters of the black hole are changing in time have been proposed. A comprehensive discussion of this subject can be found in \cite{Ashtekar:2004cn}. For our purpose, the most convenient approach was developed in \cite{Booth:2003ji,Booth:2006bn}, called the method of ``slowly evolving horizons.'' The basic idea of this method is that when the evolution of the horizon is a slow process, controlled by a small parameter $\epsilon$, one can use a future outer trapping surface as a black hole boundary. For a string with tension $\mu_s$ this parameter $\epsilon$ is small when $\mu_s\ll 1$. The future outer trapping horizon is a three-dimensional surface which may be foliated by two dimensional outer trapping surfaces. In the presence of fluxes obeying energy conditions it is a spacelike surface, which for small $\epsilon$ is close to the null one. It was demonstrated in \cite{Booth:2003ji,Booth:2006bn} that---provided some natural conditions are satisfied---one can define the generators on the slowly evolving horizons and construct generators of ``approximate symmetry'' transformations. Moreover, it was also shown that in the absence of gravitational waves the change of the energy (mass) and angular momentum of the slowly evolving black hole agrees with the expressions (\ref{EEE})--(\ref{JJJ}) obtained in the test field approximation, at least in the leading order of the smallness parameter.

As a result of the angular momentum flux through the horizon the rotation of the black holes is slowed down and its angular momentum decreases according to
\be
{J}_{(i)}={J}_{(i)}^0 e^{-v/v_i}\hh v_i=\frac{2M}{D-2} \frac{1}{ \mu_s \left( \mu_i^0 \right){}^2} \, .
\ee
This result is in agreement with our previous work on the special case of the 4D Kerr metric \cite{Boos:2017pyd}.

\section{Asymptotic torque}
Let us now demonstrate that the change of the black hole parameters \eqref{MJINF}, can be explained as the result of a torque produced by an external force that keeps the string segment at rest at infinity.

\subsection{Calculation of the asymptotic torque}

In the four-dimensional case it is straightforward to interpret the angular momentum extraction $\dot{J}^\infty$ as a result of the action of a torque on the black hole via the string \cite{Boos:2017pyd}. We prove now a similar result for the higher-dimensional case. The main difference is that in higher dimensions both angular momentum and torque are 2-forms, and only in three spatial dimensions they can be identified with 3-vectors and axial 3-vectors, respectively.

At this point it is useful to introduce the Cartesian coordinates
\begin{align}
x_i&=r \mu_i \sin\phi\, , \quad y_i=r \mu_i \cos\phi\, , \quad z=(1-\epsilon)r\mu_0 \, , \nonumber \\
i&=1,\dots,m\equiv n-1+\epsilon \, ,
\end{align}
and write the asymptotic Myers--Perry metric in the form
\begin{align}
\dd s^2 = -\dd t^2 + \sum\limits_{i=1}^m \left( \dd x_i^2 + \dd y_i^2 \right) + \left(1-\epsilon \right) \dd z^2 \, .
\end{align}
This is of course simply $D$-dimensional Minkowski space. The coordinates $x{}_i$ and $y{}_i$ are chosen such that they span the plane of rotation corresponding to the rotation parameter $a_i$. The coordinate $z$ only exists in even spacetime dimensions with $\epsilon=0$.

We denote by $\ts{e}_i$, $\hat{\ts{e}}_i$ and $\ts{e}_0$ unit vectors in the direction of the $x_i$, $y_i$ and $z$ coordinate lines, respectively. For the corresponding dual unit forms we shall use the notations $\ts{\omega}^i$, $\hat{\ts{\omega}}^i$, and $\ts{\omega}^0$. Then, the angular momentum parameters of the Myers--Perry solution can be identified with components of the following 2-form \cite{Myers:1986un}:
\begin{align}
\ts{J} = {2M\over D-2} \sum\limits_{i=1}^m  a_i \ts{\omega}_i \wedge \hat{\ts{\omega}}_i \, , \label{eq:angular-momentum}
\end{align}
where $M$ is the mass.

Let us consider a principal Killing string configuration at time $T$. Such a  string is specified by its asymptotic parameters $(\mu^0_i, \phi^0_i, (1-\epsilon)\mu_0^0 )$. In Cartesian coordinates these equations determine a straight line $L_0$ that we shall call \emph{fiducial}. Consider a sphere $S_R$ of large radius $R$
\be
R^2\equiv \sum\limits_{i=1}^m (x_i^2+y_i^2)+(1-\epsilon) z^2=\mbox{const}\, .
\ee
A unit vector $\ts{n}_R$ which is orthogonal to $S_R$ is
\be
\ts{n}={1\over R}\Big[\sum\limits_{i=1}^m (x_i \ts{e}_i+y_i \hat{\ts{e}}_i)+(1-\epsilon) z \ts{e}_0\Big]\, .
\ee

The fiducial line $L_0$ crosses  $S_R$ at a point $A_0$. By a rigid rotation it is possible to put $\phi_0=0$. From now on we use this choice of the coordinates, in which the point $A_0$ has coordinates $(x_i^0=0,y_i^0,z^0)$.
The asymptotic form of the string equation \eqref{phiass},
\be
\phi_i\sim {a_i\over r} \, ,
\ee
implies that in the asymptotic domain the string coincides with a straight line parallel to the fiducial line $L_0$ and separated from it by a finite distance
\be
\ell=\Big[ \sum \limits_{i=1}^m (a_i \mu_i^0)^2 \Big]^{\tfrac12}\, .
\ee
Thus, it crosses the sphere $S_R$ at a point slightly different from $A_0$. Such a principal Killing string segment with parameters $\mu^0_i$ will experience a relative displacement described by the impact vector\footnote{This displacement is closely related to the impact parameter of the in-coming principal null geodesic.}
\begin{align}\n{dmx}
\ts{\delta} = \sum\limits_{i=1}^m a_i \mu^0_i \ts{e}_i \, .
\end{align}

If one cuts the string at the point where it crosses $S_R$, then---in order to keep it at rest---one needs to apply the force $\mu_s$ in the radial direction $\ts{n}_R$. In the vector of this force is
\begin{align}
\ts{F} = \mu_s \ts{n}_R = \mu_s \Big[ \sum\limits_{i=1}^m \mu_i^0 \hat{\ts{e}}_i + (1-\epsilon)\mu_0^0 \ts{e}_0 \Big] .
\end{align}
This force $\ts{F}$ then acts at the displacement $\ts{\delta}$ from the position $A_0$ of the fiducial line, given by \eqref{dmx}, and creates the asymptotic torque 2-form $\ts{\tau}$. Let us denote by $\tilde{\ts{F}}$ and $\tilde{\ts{\delta}}$ the 1-forms dual to the vectors $\ts{F}$ and $\ts{\delta}$. Then, the torque 2-form is given by
\begin{align}
\label{eq:torque-2-form}
\ts{\tau} &\equiv \tilde{\ts{\delta}} \wedge \tilde{\ts{F}} = \mu_s \bigg[ \sum\limits_{i,j=1}^m a_i \mu^0_i \mu^0_j\, \ts{\omega}_i \wedge \hat{\ts{\omega}}_j \nonumber \\
&\hspace{65pt}+ (1-\epsilon)\sum\limits_{i=1}^m a_i \mu_i^0 \mu_0^0 \, \ts{\omega}_i \wedge \ts{\omega}_0 \bigg] \, .
\end{align}
Note that the impact vector, the force, and the torque 2-form remain finite in the limit $R\to \infty$.

Let us remark that $\ts{J}$ encompasses the ``internal'' angular momentum (spin) of the black hole. In order to describe a complete evolution of the black hole one needs also to describe its behavior in the surrounding bulk spacetime, that is, one needs to specify its position and velocity: the action of external forces on the black hole results in the change of its momentum and the value and orientation of its angular momentum in the bulk space.

There are two observations at this point:
\begin{itemize}
\item[(i)] If only a single string segment is attached to the black hole, the force $\ts{F} \not=0$ will accelerate the black hole.
\item[(ii)] The non-vanishing torque \eqref{eq:torque-2-form} will induce a change of the overall angular momentum of the black hole according to $\dot{\ts{J}} = \ts{\tau}$. This is because the torque 2-form has components that lie within the $x_iy_i$-planes, which will lead to a change of the angular momentum along these eigendirections of $\ts{J}$ (let us call these components ``in-plane''), but also off-diagonal terms that lie in $x_iy_j$-planes and $x_ix_j$-planes (for $i \not= j$). These contributions will lead to a time-dependent precession of $\ts{J}$ outside of its eigenplanes, hence we refer to them as ``out-of-plane.''
\end{itemize}
For any mechanical system, force and torque are additive quantities. We can therefore try to add additional string segments with suitable parameters that will counterbalance the force $\ts{F}$ and torque $\ts{\tau}$ that is caused by just one single string segment. We demonstrate now that by carefully balancing all forces and torques one can obtain  a quasi-stationary system, where the only change will be adiabatic decrease of the value of components of angular  momentum of the black hole. This somewhat resembles the construction of gearboxes in automobiles. In what follows, we will give an explicit example how to construct such a ``black hole gearbox.''

\subsection{Torque alignment procedure: constructing a black hole gearbox}
It is useful to write the displacement $\ts{\delta}$ and force $\ts{F}$ for a single string segment as
\begin{align}
\ts{\delta} &= \sum\limits_{i=\epsilon}^m \ts{\delta}_i \, , \quad \ts{F} = \sum\limits_{i=\epsilon}^m \ts{F}_i \, ,
\end{align}
where we have chosen the notation $\ts{\delta}_0 = 0$ and $\ts{F}_{0} = (1-\epsilon) \ts{F}_z$ to accommodate even and odd dimensions in one expression. Then we can write the torque of a single string segment, call it ${}^{(1)}\ts{\tau}$, as follows:
\begin{align}
{}^{(1)}\ts{\tau} &= \sum\limits_{i,j=\epsilon}^m \tilde{\ts{\delta}}_i \wedge \tilde{\ts{F}}_j = \sum\limits_{i=j} \tilde{\ts{\delta}}_i \wedge \tilde{\ts{F}}_i + \sum\limits_{i\not=j} \tilde{\ts{\delta}}_i \wedge \tilde{\ts{F}}_j \, .
\end{align}
The first term is already in-plane, but the second term is not. We can think of it as the following matrix:
\begin{align}
&\quad \sum\limits_{i\not=j} \tilde{\ts{\delta}}_i \wedge \tilde{\ts{F}}_j = \begin{pmatrix} 0 & 0 \\ \vec{T} & M \end{pmatrix} \\
&= \begin{pmatrix} 0 & \hspace{10pt} & 0 & 0 & \dots & 0 \\[10pt]
\tilde{\ts{\delta}}_1 \wedge \tilde{\ts{F}}_0 && 0 & \tilde{\ts{\delta}}_1 \wedge \tilde{\ts{F}}_2 & \dots & \tilde{\ts{\delta}}_1 \wedge \tilde{\ts{F}}_m \\
\tilde{\ts{\delta}}_2 \wedge \tilde{\ts{F}}_0 && \tilde{\ts{\delta}}_2 \wedge \tilde{\ts{F}}_1 & 0 & \dots & \tilde{\ts{\delta}}_2 \wedge \tilde{\ts{F}}_m \\
\vdots && \vdots & \vdots & \ddots & \vdots \\
\tilde{\ts{\delta}}_m \wedge \tilde{\ts{F}}_0 && \tilde{\ts{\delta}}_m \wedge \tilde{\ts{F}}_1 & \tilde{\ts{\delta}}_m \wedge \tilde{\ts{F}}_2 & \dots & 0 \end{pmatrix} \, . \nonumber
\end{align}
The matrix $M$ and the column vector $\vec{T}$ parametrize the out-of-plane torque. For $\epsilon=1$ this column vector $\vec{T}$ and the extra row with zeros are absent, so that the out-of-plane torque reduces to a square $m\times m$ matrix $M$.

Let us now introduce the following set of discrete point transformations:
\begin{align}
\begin{split}
\label{eq:discrete-transformations}
I_0 &: z \rightarrow -z \, , \quad I_j : (x_j, y_j) \rightarrow (-x_j, -y_j) \, , \\
I &: (x_1, y_1, \dots, z) \rightarrow (-x_1, -y_1, \dots, -z) \, .
\end{split}
\end{align}
They correspond to an inversion along the $z$-axis, rotations of $\pi$ in the $x_j y_j$-planes, and a total inversion, respectively. It is easy to see that string segments parametrized by $(x_i^0, y_i^0, z^0)$ and $I_j (x_i^0, y_i^0, z^0)$ (for some fixed $j$) have the same in-plane torque. However, their out-of-plane torque matrix $M$ differs by a sign in the $j$-th row and $j$-th column because these components only contain either $x^0_j$ or $y^0_j$. Adding up these torques hence eliminates the $j$-th row and $j$-th column from the total torque.

\begin{figure}
  \centering
  \includegraphics[width=0.3\textwidth]{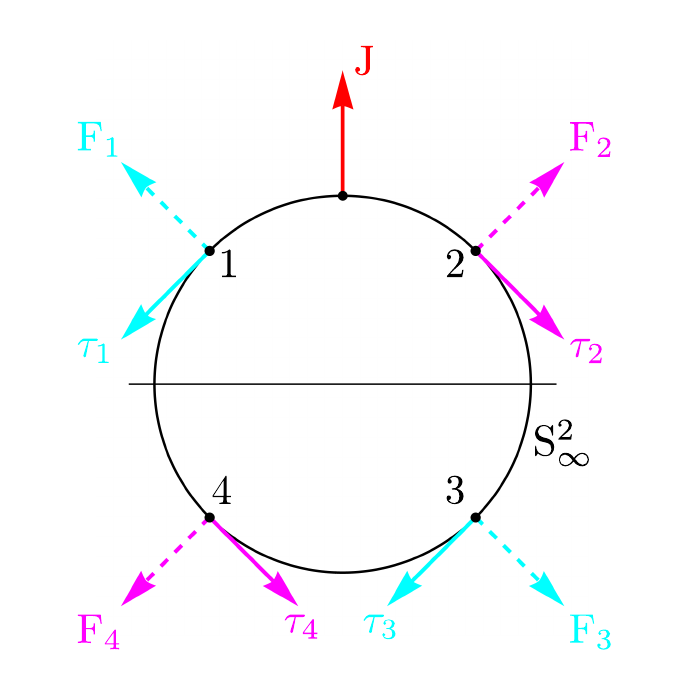}
  \caption{Torque alignment procedure for the 4D case: The angular momentum 2-form $\ts{J} = Ma \dd x \wedge \dd y$ corresponds to the 3-vector $\partial_z$ pointing along the $z$-axis. For a string with parameters of point 1 on $S^2_\infty$, first create a second string at point 2 to compensate the out-of-plane component of the torque. Then, add the two strings at points 3 and 4 to cancel the net force while keeping the net torque aligned to $\ts{J}$.}
  \label{fig:gear-box-4d}
\end{figure}

One can now repeat this step for all planes $j=1,\dots,m$ and thereby set $M=0$, doubling the number of string segments in each step. The vector $\vec{T}$ is non-vanishing for even spacetime dimensions. It can be set to zero by inserting a $z$-inverted copy of each string segment, thereby doubling the number of string segments one more time. We obtain
\begin{align}
{}^{(2^{n-1})}\ts{\tau} = \Big[ \prod\limits_{i=\epsilon}^m \left( \mathbb{1} + I_i \right) \Big] {}^{(1)}\ts{\tau} \, ,
\end{align}
which amounts to the net torque of $2^{m-\epsilon} = 2^{n-1}$ principal Killing string segments, which is in-plane with $\ts{J}$ such that $\ts{J}$ will not pick up any non-diagonal terms. We can finally set the net force to zero by doubling the number of principal Killing string segments one last time via adding a copy of each string segment at the totally inverted position such that
\begin{align}
{}^{(2^n)}\ts{F} &= (\mathbb{1} + I) {}^{(2^{n-1})} \ts{F} = 0 \, \\
{}^{(2^n)}\ts{\tau} &= (\mathbb{1} + I) {}^{(2^{n-1})} \ts{\tau} \, .
\end{align}
In general, there is a total number of $2^n$ principal Killing string segments required to perform this torque alignment procedure. See Fig.~\ref{fig:gear-box-4d} for an explicit example of this procedure for the four-dimensional case. Since all of these string segments extend from the black hole horizon to spatial infinity, and all of the string segments (via the last $I$-transformation) have an antipodal counterpart on the other side of the black hole, one can equivalently think of $2^{n-1}$ infinite strings that pierce the central Myers--Perry black hole and extend from spatial infinity to spatial infinity.

The above procedure has a group-theoretical interpretation. The spatial isometry group of the Myers--Perry spacetime is (for all $a_i$ non-vanishing and $a_i \not= a_j$)
\begin{align}
\mathcal{G} = \mathbb{Z}^{1-\epsilon} \times \text{SO}(2)^m \, ,
\end{align}
where $\mathbb{Z}$ denotes the reflection symmetry with respect to the $z$-axis in even spacetime dimensions ($\epsilon=0$), and $\text{SO}(2)^m$ encodes the axisymmetry in each $x_iy_i$-plane with $i=1,\dots,m$. In the presence of a single principal Killing string segment, $\mathcal{G}$ gets broken to the trivial group, and no isometries remain.

The construction of the black hole gearbox, however, restores the isometries somewhat. The spacetime of the Myers--Perry black hole with $2^n$ principal Killing string segements attached has the discrete isometry group
\begin{align}
\mathcal{G}' = I_0 \times (I_j)^m \cong \mathbb{Z}^{1-\epsilon} \times \mathbb{Z}^{m} ,
\end{align}
where we used the notation of Eq.~\eqref{eq:discrete-transformations}. Moreover, $\mathbb{Z}^{1-\epsilon}$ again denotes a discrete symmetry with respect to reflections about the $z$-axis in even spacetime dimensions ($\epsilon=0$), and $\mathbb{Z} \subset \text{SO}(2)$ corresponds to the reflection symmetry under the discrete subgroup of two-dimensional rotations. The above seems to mimic the emergence of approximate symmetries in the context of small perturbations \cite{Booth:2003ji,Booth:2006bn}, but further work is required to determine whether there is a deeper connection or not.

\section{Discussion}
\label{sec:conclusions}

In a spacetime which admits a principal tensor $\ts{h}$ there exists a rich geometrical structure generated by this object \cite{Frolov:2017kze}. As a result of the existence of a set of explicit and hidden symmetries (Killing tower) generated by the principal tensor, the geodesic equations are completely integrable and several physical field equations allow complete separation of variables. A natural question is: Can these symmetries be used to solve equations for more complicated physical objects? In particular, this question is highly interesting for the case of strings propagating in a spacetime with the principal tensor. Unfortunately, in the general case of dynamical strings the answer seems to be negative. However, there exists a wide class of string configurations for which the equations are proven to be completely integrable \cite{Kubiznak:2007ca}. The characteristic property of such strings is that their worldsheet is tangent to the primary Killing vector of $\ts{h}$. Within this class of strings there exist special solutions which we called principal Killing strings. In this paper we studied properties of these string solutions in the general off-shell Kerr--NUT--(A)dS spacetime with an arbitrary number of dimensions.

First, we demonstrated that in a spacetime with the principal tensor $\ts{h}$ a two-dimensional surface $\Sigma_{\pm}$ tangent to the primary Killing vector and the principal null vector $\ts{l}_{\pm}$ is a minimal surface. Thus, this surface is a solution of the Nambu--Goto equations and represents a stationary string in the Kerr--NUT--(A)dS spacetime. In particular, in the presence of the future event horizon $\mathcal{H}_+$, the surface $\Sigma_-$ represents a string that is regular at $\mathcal{H}_+$ in incoming null coordinates. The strings $\Sigma_+$ have a similar property for the past horizon $\mathcal{H}_-$.

Stationary string configurations in the class of Kerr--NUT--(A)dS geometries have been discussed in the context of complete integrability \cite{Kubiznak:2007ca}. This is a powerful technique, but the results are often of a parametric form which makes it difficult to extract an explicit parametrization. Using our approach based on the principal tensor, however, we obtained an explicit solution for principal Killing strings in the off-shell metric, see Sec.~\ref{sec:pks}, Eq.~\eqref{eq:pks-main-result}. Moreover, we also found a coordinate transformation from the canonical coordinates, connected with the principal tensor, to new coordinates in which the string configuration is straightened. This allowed us to obtain a simple expression for the stress-energy tensor of the test string in the off-shell metric.

To illustrate these general results we discussed the special case of the Myers--Perry metric. In Sec.~\ref{sec:myers-perry}, we explicitly showed that for such metrics our results reproduce the results obtained earlier \cite{Frolov:2004qw}. It should be emphasized that the paper \cite{Frolov:2004qw} was written before the existence of the principal tensor for Myers--Perry metric was discovered in \cite{Frolov:2007nt}, and hence our new approach represents a significantly streamlined calculation.

In addition, we proved a convenient version of the energy and angular momentum conservation laws which allow one to demonstrate the equality of the energy and angular momentum fluxes through the horizon and similar fluxes through special chosen surfaces of constant radius $r$. We also calculated directly the energy and angular momentum fluxes at infinity and demonstrated the consistency of these results with the calculations at the horizon. Another new result is the relation of the change of the parameters of the rotating black hole pierced by a string to the torque produced at infinity by forces keeping the string stationary. To that end, we discussed special configurations of a set of strings such that their net action on the black hole is a reduction of its rotation parameters.

It should be emphasized that besides these ``internal'' degrees of freedom such as the mass and the angular momenta, the black hole is characterized by its ``external'' degrees: its position and velocity (momentum) in space, as well as the orientation of its spin. In general, the action of a single string segment on the black hole does not just change its angular momenta: the string segment's force and torque also have components that result in the change of the momentum of the black hole and its spin orientation (precession). However, as we demonstrated, these effects can be counterbalanced by attaching several other string segments to the black hole. First of all, the net force acting on the black hole vanishes for two pieces of string segments piercing the black hole in two antipodal points. Such a configuration can naturally represent a single infinite string captured by a black hole. Moreover, one can find a combination of several strings for which the net torque effect slows down the spin of the black hole, while the other torque components, responsible for the change of the spin orientation, vanish identically.

\section*{Acknowledgments}
We thank our anonymous referees for their comments. J.B.\ is grateful for a Vanier Canada Graduate Scholarship administered by the Natural Sciences and Engineering Research Council of Canada as well as for the Golden Bell Jar Graduate Scholarship in Physics by the University of Alberta.
V.F.\ thanks the Natural Sciences and Engineering Research Council of Canada and the Killam Trust for financial support.

\bibliography{STRINGY_BHS_4}{}

\end{document}